\pdfoutput=1
\documentclass[12pt]{article}
\usepackage[margin=.99in]{geometry}

\usepackage{amsmath,dsfont}
\usepackage{amssymb}
\usepackage{graphicx,color}
\usepackage{slashed,cite,xcolor}
\usepackage{hyperref}

\usepackage{ifpdf}

\newcommand{\be}{\begin{equation}}
\newcommand{\ee}{\end{equation}}
\newcommand{\ba}{\begin{eqnarray}}
\newcommand{\ea}{\end{eqnarray}}

\title{{\bf \bigskip \Large{The asymptotic dS Swampland Conjecture -\\ a simplified derivation and a potential loophole}}\bigskip}

\author{Arthur Hebecker$^a$ and Timm Wrase$^b$\bigskip\bigskip\\
\small $^a$Institute for Theoretical Physics, University of Heidelberg, \\ \small Philosophenweg 19, D-69120 Heidelberg, Germany\bigskip\\
\small $^b$Institute for Theoretical Physics, TU Wien,\\ \small Wiedner Hauptstrasse 8-10/136, A-1040 Vienna, Austria\bigskip
}

\date{}

\begin{document}

\maketitle

\abstract{
\noindent Recently, arguments for a refined de Sitter conjecture were put forward in \cite{Ooguri:2018wrx}. Using the large distance conjecture of \cite{Ooguri:2006in}, the authors provide evidence for this dS conjecture in asymptotic regimes of field space, where the parametric control of string theory becomes arbitrarily good. Their main tool is Bousso's covariant entropy bound \cite{Bousso:1999xy}. Here, we describe a simpler way to reach a similar conclusion. The argument is based on the fact that the cutoff of an effective theory with gravity decreases as the number of species grows. We then discuss a loophole in this argument and its possible counterpart in the assumptions underlying the entropy-based derivation. The idea is to consider potentials which, while they remain below an exponentially falling bound, have small oscillations leading locally to relatively flat regions or even to an infinite series of dS minima.
}

\thispagestyle{empty}


\newpage


\section{Introduction}
Following the proposal of the so called dS swampland conjecture~\cite{Obied:2018sgi} (see \cite{Brennan:2017rbf, Danielsson:2018ztv} for related reviews), a constructive debate concerning dS models in string theory has been revived. Moreover, the related phenomenology of dark energy and inflation has received much attention, see e.g.~\cite{Agrawal:2018own, Andriot:2018wzk, Dvali:2018fqu, Achucarro:2018vey, Garg:2018reu, Lehners:2018vgi, Kehagias:2018uem, Dias:2018ngv, Denef:2018etk, Colgain:2018wgk, Paban:2018ole, Roupec:2018mbn, Andriot:2018ept, Matsui:2018bsy, Ben-Dayan:2018mhe, Heisenberg:2018yae, Damian:2018tlf, Conlon:2018eyr, Kinney:2018nny, Dasgupta:2018rtp, Cicoli:2018kdo, Kachru:2018aqn, Akrami:2018ylq, Murayama:2018lie, Marsh:2018kub, Choi:2018rze, DAmico:2018mnx, Han:2018yrk, Moritz:2018ani, Halverson:2018cio, Anguelova:2018vyr, Lin:2018kjm, Hamaguchi:2018vtv, Ashoorioon:2018sqb, Wang:2018kly, Fukuda:2018haz}. Clearly, the realization of dS minima and of inflation in string theory is a difficult and important subject and such a revived attention is very desirable.
An important point in our context is the suggestion of potential counter-examples to the original dS swampland conjecture in \cite{Denef:2018etk, Conlon:2018eyr, Cicoli:2018kdo, Murayama:2018lie, Choi:2018rze, Hamaguchi:2018vtv}. The refined version of the dS swampland conjecture advertised in \cite{Ooguri:2018wrx}, which is very close to that of~\cite{Garg:2018reu}, does not have these issues (see \cite{Dvali:2018fqu, Andriot:2018wzk} for earlier suggestions for refinement). Very roughly, the statement of the refined conjecture is that $|V'|/V>{\cal O}(1)$ whenever the potential is sufficiently strongly convex. The asymptotic form claims only that this holds as one moves to infinity in field space.

In \cite{Ooguri:2018wrx} the authors also point out an interesting connection between their refined dS swampland conjecture and the swampland distance conjecture \cite{Ooguri:2006in, Cecotti:2015wqa, Palti:2015xra, Baume:2016psm, Klaewer:2016kiy, Valenzuela:2016yny, Blumenhagen:2017cxt, Palti:2017elp, Lust:2017wrl, Hebecker:2017lxm, Cicoli:2018tcq, Grimm:2018ohb, Heidenreich:2018kpg, Blumenhagen:2018nts, Lee:2018urn}. They do so by using  Bousso's covariant entropy bound \cite{Bousso:1999xy} and thereby provide support for their refined dS swampland conjecture in the asymptotic regime, which is parametrically controlled in a strict, mathematical sense. This is a generalization of the old Dine-Seiberg argument \cite{Dine:1985he}.
This result of \cite{Ooguri:2018wrx} is also consistent with the well-known fact (recalled e.g. in the first paragraph of \cite{Denef:2018etk}) that dS vacua must live in a regime where corrections cannot be made arbitrarily small since a single leading (non-constant) term in the scalar potential always gives rise to a runaway. This is also consistent with existing constructions of dS vacua like the KKLT \cite{Kachru:2003aw} and LVS scenario \cite{Balasubramanian:2005zx}, where several terms compete with each other to give rise to dS solutions. 

While the authors of \cite{Ooguri:2018wrx} also consider the possible application of their refined dS swampland conjecture to the entire moduli space, we will focus here only on the asymptotic region, i.e., on the limit where we approach the boundary of field space. We discuss a simple argument, based on the species bound~\cite{Veneziano:2001ah, ArkaniHamed:2005yv, Dvali:2007hz, Dvali:2007wp}, why one expects that the potential decays exponentially in any such direction.\footnote{We have been informed that the authors of~\cite{Ooguri:2018wrx} were aware of the restrictions imposed by the species bound in their context. The connection between the species bound and the Dine-Seiberg argument has also been previously pointed out by Matt Reece~\cite{talk-madrid}. For earlier discussions of the species and entropy bounds in the context of quasi-de-Sitter solutions see e.g.~\cite{Ashoorioon:2011aa,Conlon:2012tz, Kaloper:2015jcz}.}
While this asymptotic behavior seems generically consistent with the refined dS swampland conjecture, we point out a simple loophole based on an exponentially bounded potential that nevertheless has dS vacua. We then analyze how this loophole might show up in the assumptions of the entropy-based derivation. It would be interesting to check whether such a potential can or cannot arise in string theory.

\section{A simplified derivation}
We now use the distance conjecture \cite{Ooguri:2006in, Baume:2016psm, Klaewer:2016kiy} and the reduced cutoff in the presence of a large number of species \cite{Veneziano:2001ah, ArkaniHamed:2005yv, Dvali:2007hz, Dvali:2007wp} to show that the scalar potential should decay exponentially when we approach the boundary in field space.\footnote{For other applications of towers of light states in related contexts see, e.g.,~\cite{Heidenreich:2018kpg, Andriolo:2018lvp}.} Concretely, we argue that for canonical $\phi$ the scalar potential has to satisfy $V < e^{-a \phi}$ for $\phi\rightarrow \infty$, where $a$ is an order one parameter.

Let us focus on a single tower of light states and assume equal spacing, as in the case of KK or string winding modes. In other words, we assume that the masses are given by $n \cdot m$, where $n\in \mathbb{N}$ and $m\equiv\, M_P  e^{-b \phi}$ with some number $b$ that is expected to be order one according to the distance conjecture. The number of particles with a mass below the cutoff is given by $N \approx \Lambda/m$. Next, similarly to \cite{Grimm:2018ohb}, we combine this with the well-known result \cite{Dvali:2007wp} that the cutoff of an effective gravitational theory is reduced in the presence of a large number of species, $\Lambda \approx M_P /\sqrt{N}$. Eliminating $N$ from these two relations, we find the highest value of $\Lambda$ for which the theory with this tower of states remains perturbatively controlled:
\be
\Lambda \approx (M_P^2 m)^{\frac13} \approx M_P e^{-\frac{b\phi}{3}}\,.
\ee 
We therefore find that the cutoff $\Lambda$ decreases exponentially, with the order one pre-factor $b/3$. In the case of multiple towers or, more precisely, a higher-dimensional lattice of light states, the prefactor changes but the crucial exponential behavior will remain.

Now let us allow for a positive scalar potential $V(\phi)$. Such a potential will in general (with details depending on its steepness and the overall dynamics) introduce a curvature scale $\sim H$, with $H$ defined by $3H^2M_P^2=V$. To have perturbative control, we need an effective field theory regime at a scale above $H$, i.e., we need $\Lambda\gtrsim H$. In other words, 
\be
M_P^2e^{-2b\phi/3}\gtrsim V(\phi)/M_P^2\,.
\ee
which is our desired exponential bound on the potential.

This result differs in two points from the one obtained in \cite{Ooguri:2018wrx}. Firstly, it is only an upper bound while the argument in \cite{Ooguri:2018wrx} gives the actual behavior of the potential. Secondly, it relies only on the reduced cutoff and the distance conjecture. By contrast, the derivation in \cite{Ooguri:2018wrx} requires an apparent horizon. This constrains in particular the second derivative of the scalar potential, which hence motivated the refined dS swampland conjecture in \cite{Ooguri:2018wrx}.

\section{A potential loophole}
We have provided evidence, in an arguably simpler way than that of \cite{Ooguri:2018wrx}, for the exponential decay of the scalar potential. Let us now assume that indeed $V(\phi) < e^{-a \phi}$ as one approaches the boundary of field space, $\phi \rightarrow \infty$. One would like to further conclude that $|V'|/V\gtrsim a$ since the potential decays with an exponent $-a$ or faster. While we do expect this behavior in string theory, it is unfortunately not so easy to derive this last result in a strict sense. Indeed, consider the following simple counter-example:

If we multiply the exponentially decaying potential by an oscillating function, e.g.
\be
V(\phi)=e^{-a\phi}[A+B\cos(\alpha\phi)]\,,\label{pot}
\ee
then such a potential can have metastable dS vacua.\footnote{
See \cite{Palti:2017elp} for a different use of the idea of oscillations to establish a loophole in a Swampland-type argument.} Moreover, even imposing monotonicity as an extra ad hoc constraint, the desired conclusion cannot be reached. Given appropriately chosen values of the parameters in (\ref{pot}), one can ensure that $V$ falls monotonically but has infinitely many, arbitrarily flat regions. Such oscillations are compatible with the species-bound argument above.

They do not provide a loophole in the logic of \cite{Ooguri:2018wrx} but may rather correspond to some of the assumptions in \cite{Ooguri:2018wrx} being too strong. Specifically, the logic presented there postulates a monotonic behavior of the prefactor $n(\phi)$ in the exponential growth of species, $N(\phi)=n(\phi)\exp(c\phi)$, below a certain cutoff (which must also depend on $\phi$ and the definition of which is not a priori obvious). Then, this number $N$ of species has to saturate the dS entropy precisely.
\footnote{Let us note, independently of our main logic, that in controlled string models this would be very special. Indeed, the known asymptotic directions are large volume $(t\to \infty)$, large complex structure $(z\to \infty)$, and small $g_s$ ($g_s\to 0$). In the last case, the light tower are string excitations and stabilizing the compactification space above that scale is hard. By the Strominger-Yau-Zaslov conjecture \cite{Strominger:1996it}, going to large $z$ at fixed $t$ means going to large volume of the mirror dual. So we can focus on the latter case. But, at large volume, the light tower starts at the KK scale. Now (setting $M_P=1$) one may use the free-particle entropy relation $S\sim N^{1/4}R^{3/2}$ \cite{Ooguri:2018wrx}  for $N$ species from above $m_{KK}$. Together with $S\sim R^2$ and $R\sim 1/H$, this gives $H\sim 1/R\sim 1/\sqrt{N}$. But $1/\sqrt{N}$ is the species cutoff, implying $1/\sqrt{N}\gg m_{KK}$ and hence $H\gg m_{\rm KK}$. This differs from conventional settings where $H\ll m_{\rm Moduli}\ll m_{\rm KK}$ and it would be interesting to see how such an unusual regime can arise.}
Thus, to close the oscillations-loophole using the entropy argument, one would need a precise definition of the relevant number of species (via the cutoff or otherwise), and furthermore demonstrations that this number grows in a precisely exponential way (as above) {\it and} saturates the dS entropy precisely.  By precisley we here mean without any `wiggle room' of ${\cal O}(1)$ factors. We believe that more work is needed to understand these points and to exclude oscillations on general grounds.

Until then, it remains an interesting question whether such a scalar potential (or a variant thereof) can arise in string theory. It does not seem likely that saxion directions would receive such periodic corrections. However, one can break the shift symmetry along axionic directions~\cite{Silverstein:2008sg, McAllister:2008hb}, for example by a flux-induced $F$-term potential~\cite{Marchesano:2014mla, Blumenhagen:2014gta, Hebecker:2014eua}. The desired outcome of this was an axion-monodromy-type inflationary potential. At least in some cases, the appearance of the latter may be prevented by the backreaction of the axion on the saxion~\cite{Palti:2015xra, Baume:2016psm, Klaewer:2016kiy} (as reviewed in Sect.~5 of \cite{Blumenhagen:2018hsh}). Either way one always expects, based on the argument of the previous section, to find an asymptotically decaying potential that in this case could have oscillations. We leave it to the future to construct or rule out models with infinite directions in field space that involve the axion and saxion and give rise to such asymptotic oscillations.

\section{Conclusion}
We have discussed a simple argument for an exponential bound on scalar potentials at the boundary of moduli space: $V(\phi) < e^{-a \phi}$ for $\phi \rightarrow \infty$ and with $a\sim{\cal O}(1)$. The argument combines the appearance of many light fields according to the distance conjecture \cite{Ooguri:2006in} with the species bound \cite{Veneziano:2001ah, ArkaniHamed:2005yv, Dvali:2007hz, Dvali:2007wp}. The latter then limits the UV cutoff for $\phi\to \infty$. While this argument is rather straightforward and has probably been known to many of the experts, we find its discussion worthwhile for two reasons.

Firstly, it complements the derivation in~\cite{Ooguri:2018wrx}, where the authors argue for the actual behavior of the scalar potential: $V \sim e^{-b\phi}$. This result requires assumptions about or the knowledge of several ${\cal O}(1)$ factors in the interplay between cutoff, species number and de Sitter entropy. From the perspective of our simpler argument these assumptions correspond to the loophole of an exponentially bounded but oscillating potentials. In this sense, our note highlights a particularly non-trivial and possibly not yet finalized aspect of the entropy-based derivation of $V \sim e^{-b\phi}$.

Secondly, by suggesting a concrete way in which de Sitter or slow-roll regions for $\phi\to \infty$ can arise (e.g.~via backreacted axion-monodromy potentials), we open up the possibility of trying to challenge the asymptotic de Sitter conjecture in a constructive way. It may be an interesting exercise to attempt the realization of such scalar potentials in string theory. Alternatively, one can also prove their non-existence by establishing the assumptions on which the derivation of \cite{Ooguri:2018wrx} is based. Both of avenues provide important and feasible research projects for the future.

\vspace*{.2cm}
\noindent
{\bf Acknowledgments:} 
We would like to thank Philipp Henkenjohann, Eran Palti and Pablo Soler for useful discussions. TW is supported by an FWF grant with the number P 30265. 

\bibliographystyle{JHEP}
\bibliography{refs}

\end{document}